\documentclass[conference]{IEEEtran}
\IEEEoverridecommandlockouts

\usepackage{geometry}
\usepackage{cite}
\usepackage{floatrow}
\usepackage{amsmath,amssymb,amsfonts}

\usepackage{algorithm}
\usepackage{algpseudocode}
\usepackage{diagbox}
\usepackage{graphicx}
\usepackage{textcomp}
\usepackage{stfloats}
\usepackage{amstext}
\usepackage{mathtools}
\usepackage{bigints}
\usepackage{color}
\usepackage{xcolor}
\usepackage{cite}
\usepackage{array}
\usepackage[colorlinks, urlcolor=black, linkcolor=., citecolor=black]{hyperref}
\usepackage[caption=false,font=footnotesize]{subfig}
\usepackage{blindtext}
\usepackage{float}
\floatstyle{plaintop}
\restylefloat{table}

\usepackage{amsthm}
\theoremstyle{remark}

\newtheorem{remark}{Remark}
\newtheorem{problem}{Problem}

\definecolor{myOrange}{rgb}{1,0.5,0}

\def\BibTeX{{\rm B\kern-.05em{\sc i\kern-.025em b}\kern-.08em
    T\kern-.1667em\lower.7ex\hbox{E}\kern-.125emX}}

\begin{document}

\title{\LARGE \bf A Predictive Autonomous Decision Aid for Calibrating Human-Autonomy Reliance in Multi-Agent Task Assignment}
\author{Larkin~Heintzman and Ryan~K.~Williams
\thanks{
Larkin~Heintzman and Ryan~K.~Williams are with the Department of Electrical and Computer Engineering, Virginia Polytechnic Institute and State University, Blacksburg, VA USA, 
\mbox{E-mail: \textit{\{hlarkin3, rywilli1\}@vt.edu}}.
}%
\thanks{This work was supported by the National Science Foundation under grants CNS-1830414.}
}

\maketitle

\begin{abstract}
In this work, we develop a game-theoretic modeling of the interaction between a human operator and an autonomous decision aid when they collaborate in a multi-agent task allocation setting. In this setting, we propose a decision aid that is designed to calibrate the operator's reliance on the aid through a sequence of interactions to improve overall human-autonomy team performance. The autonomous decision aid employs a long short-term memory (LSTM) neural network for human action prediction and a Bayesian parameter filtering method to improve future interactions, resulting in an aid that can adapt to the dynamics of human reliance. The proposed method is then tested against a large set of simulated human operators from the choice prediction competition (CPC18) data set, and shown to significantly improve human-autonomy interactions when compared to a myopic decision aid that only suggests predicted human actions without an understanding of reliance.
\end{abstract}

\section{Introduction} \label{sec:introduction}                              

Recently there has been much attention given to how machines can better interact with humans \cite{music2017control,okamura2020empirical}. Indeed the idea that reliance on autonomy must be calibrated is not new \cite{lee2004trust,gao2006extending}, however with an increasingly automated world there has been more focus given to how reliance can be affected by the machine itself \cite{chen2020trust,losey2019robots}, as well as how to better link human and machine decision making \cite{burton2020systematic}. As detailed in the seminal work \cite{lee2004trust}, using automation to assist with a task can result in sub-optimal outcomes if the operator is either under \emph{or} over weighting the effectiveness of the automation.  For our running example in this paper, consider an \emph{autonomous} multi-agent search and rescue (SAR) mission \cite{Williams2020-sv,Heintzman2021-mw} where the goal is to locate a lost person by searching areas of land (referred to as sectors) and a human mission operator is using a decision aid to help assign tasks to each searching agent (unmanned aerial vehicles (UAVs) or human searchers). A decision aid in a SAR context can certainly take many forms, such as computing likely lost person trajectories \cite{hashimoto2019agent}, likelihood of survival given environment \cite{xu2008usariem}, determining ideal search tasking/plans \cite{Heintzman2020-cz,Heintzman2021-mw,Heintzman2021-qr}, and so on. The risk involved in such a scenario is quite high, thus the decision aid and operator would need to be calibrated for such an environment. If the operator were to under weight the decision aid's information then critical information may be being ignored, mitigating the effectiveness of the multi-agent team. Similarly, if the operator were \emph{over} weighting the aid's information then the indispensable first-hand experience of SAR professionals may be being inhibited. As detailed in \cite{lee2004trust}, no automated system is free from faults thus the danger of an operator becoming complacent should be considered. Readers are referred to \cite{hashimoto2019agent} and \cite{koester2008lost} for additional information on state-of-the-art SAR operations.

In this work, we consider a game-theoretic modeling of the interaction between a human operator and an autonomous decision aid (ADA) in a multi-agent task allocation setting.  In this setting, we assume the operator has ultimate control over the final tasks selected, however it is the goal of the aid to correctly calibrate the operator's reliance upon the aid and to improve long-term team performance. In order to accomplish this goal, an ADA must consider several pieces of information related to the operator's usage of the aid. Namely, the decision aid must consider both what \emph{task} the operator may select (action or selection prediction), as well as the operator's likely preference towards the aid itself (reliance indication). We approach this problem while assuming a generic multi-agent task allocation scenario, with the SAR context as inspiration. To the best of the authors' knowledge, our design of a predictive decision aid for human-autonomy interactions in a multi-agent task allocation context, with the reliance model used therein, is the first of it's kind.

\begin{figure}
    \centering
    \includegraphics[width = 0.95\linewidth]{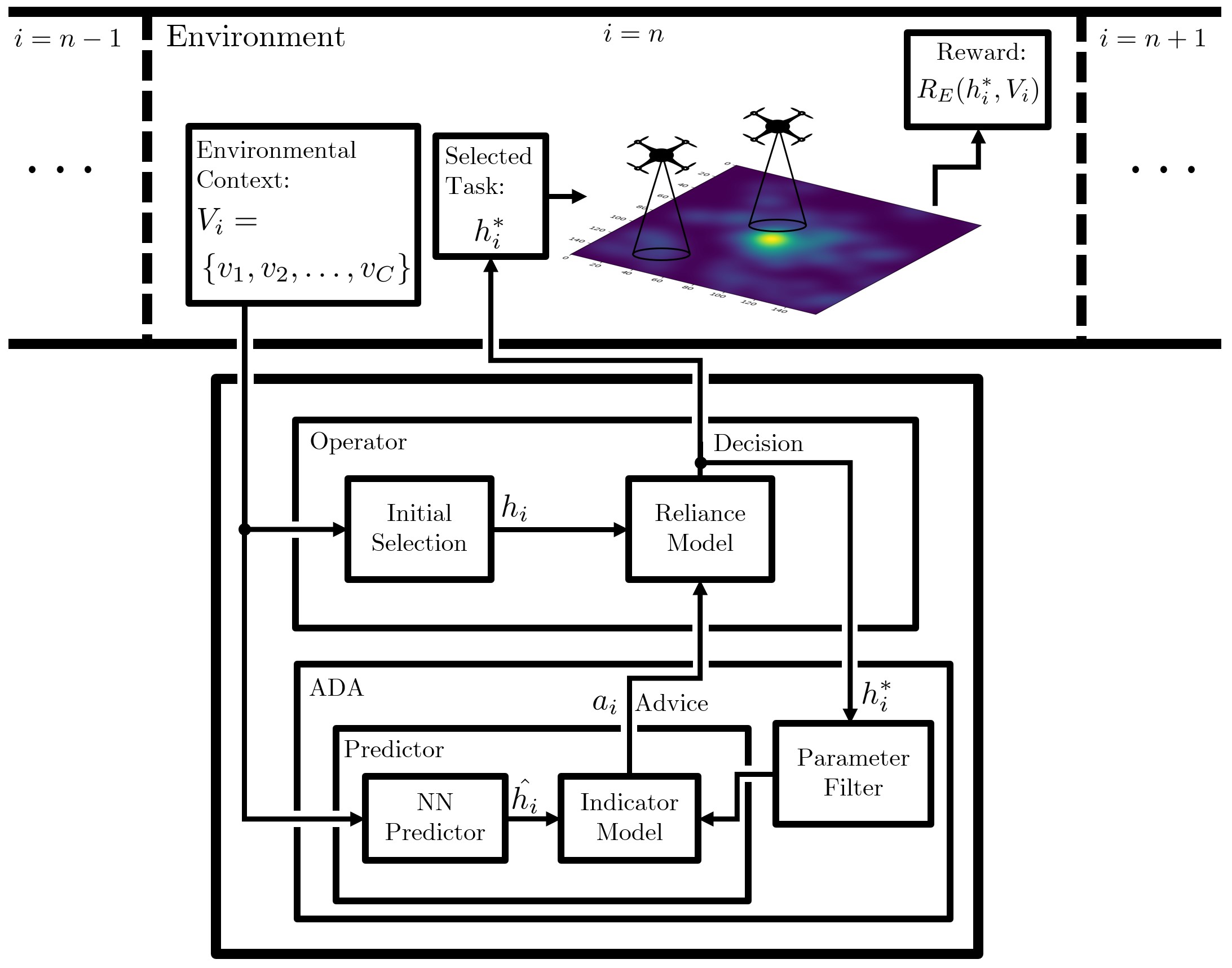}
    \caption{Showing a simplified overall system diagram for the proposed work.}
    \label{fig:system_diagram}
\end{figure}

Several recent works have considered decision aid design as it relates to trust/reliance of multi-agent systems \cite{nam2019models,wang2014human} as well as active reliance calibration \cite{okamura2020empirical,losey2019robots}. In \cite{nam2019models}, the authors investigate human trust dynamics when interacting with a semi-autonomous aerial multi-agent swarm carrying out various exploration tasks. An earlier work \cite{wang2014human} also considers a similar problem, though with underwater vehicles, and a simple switching control method. In \cite{nam2019models}, the autonomy and human operator are working in concert to drive the swarm to various locations in an unknown environment, while the operator uses a sliding scale to give trust evaluations. Similar to our work, \cite{nam2019models} uses an adaptive predictor to estimate trust dynamics which is updated as new information becomes available. However, the predictor has input requirements specific to swarm dynamics, such as agent heading variance and convex hull area, where as we design a more general LSTM-based predictive aid that requires only prior interaction data.

In \cite{okamura2020empirical}, the authors propose to adjust a human operator's trust via so-called trust calibration cues (TCCs), which are designed to notify the operator of under/over reliance during a simulated UAV piloting task. The main focus of \cite{okamura2020empirical} is in the design of a trust calibration AI which controls the frequency and type of calibration cues given to the operator. We differ from this work in our problem setting of discrete sequential interactions between operator and aid, and our purposeful limiting of aid suggestions.

Also dealing in multi-agent trust calibration, the authors of \cite{chen2020trust} consider interaction between a user and a robot working together on a table-clearing task. The robot employs a trust POMDP to inform its decisions with a goal of improving team performance in the long term. The human user exists in a supervisorial role and only intervenes when they feel the robot may not succeed in its chosen task, which is then used to inform the trust POMDP model. The resultant robot behaviors primarily attempted to build trust, though interestingly the robot would sometimes intentionally fail to correct an inappropriate level of trust. However, whereas \cite{chen2020trust} requires a numerical evaluation of trust directly from each user, we seek to create a reliance calibration method that does not require such evaluations. Also we pose the interaction as both the human and ADA having input to cooperative problem, in our case multi-agent task allocation.

\section{Preliminaries and Problem Formulation} \label{sec:prelims}
\subsection{Game Structure} \label{sub:game_structure}

We model the interaction between the human operator and the autonomy (ADA) in a game-theoretic sense. Taking nomenclature from the field of game theory, the interaction is modeled as a two player sequential imperfect information game. The result of an interaction between human and autonomy is a task for a single agent $s \in S_i$, where $S_i$ is the task set available to the multi-agent team at iteration $i$, to execute given some reward associated with the selection.  In this work, we will consider tasks related to our SAR example, however it is important to point out the generality of a task allocation model.  Indeed, the allocation of tasks to multi-agent teams can be mapped to a wealth of both theoretical efforts in multi-agent systems \cite{Williams2017-ca,liu2019submodular}, and to application-oriented domains such as target tracking \cite{Sung2020-tl,Liu2021-gw,Sung2018-mm}, agricultural monitoring \cite{liu2020data,liu2020coupled,liu2018optimal,liu2020monitoring}, etc.

Upon selecting a task, $s$, an additive reward is gained based on $R_E(s, V_i) \in \mathbb{R}$ which is a stochastic function of the selected task and the current environment state. Here, $V_i \in \mathbb{R}^L$ is an environmental context vector. Going back to our running multi-agent SAR example, the context vector would represent the state of the search such as lost person location probabilities \cite{koester2008lost}, and the reward would represent information gain from executing the selected task potentially affecting later decisions. The game is repeated for a known number of iterations, $K$.

The order of interaction between operator and ADA proceeds as follows:
\begin{itemize}
    \item The operator decides which task to select without ADA input, label this task $h_i$.
    \item The ADA has the opportunity to make a suggestion of a particular task to select, potentially different from the task previously selected by the operator, label the suggestion $a_i$.
    \item The human operator considers the given suggestion and decides whether to keep their original selection or to switch and agree with the ADA, resulting in task $h_i^*$.
    \item The selected task is carried out according to the operator's decision and reward $R_E(h_i^*, V_i) $ is gained.
\end{itemize}
For a visual model of the proposed work, consider Figure~\ref{fig:system_diagram} where the process begins with the context vector $V_i$ which informs both the predictor (Section~\ref{sub:predictor}) as well as the operator. Once the initial selection, $h_i$, and prediction, $\hat{h}_i$, are made, the indicator model (Section~\ref{sub:indicator_model}) is used to determine the suggestion provided, $a_i$, which is then returned to the operator. The final task, $h^*_i$, is selected based on the reliance decision of the operator which in turn is returned to the parameter filter to improve the indicator model.

\subsection{Operator Reliance Model} \label{sub:reliance}

We model the operator as using a stochastic reliance model that determines ADA reliance during an interaction. Reliance upon, or \emph{trust of}, an autonomous aid can change quickly and responds to certain conditions, thus the reliance model used in this work aims to capture the relevant behaviors \cite{akash2018classification, dubois2020adaptive}. We base our operator reliance model on the well known decision field theory (DFT) models from \cite{gao2006extending,dubois2020adaptive}. The definition of \emph{preference}, which directly determines reliance, is given as:
\begin{equation} \label{eq:baseReliance}
\begin{aligned}
    P(n) &= (1-s)P(n-1) + sB_{C}(n) + \epsilon(n)\\
\end{aligned}
\end{equation}
where $P(n) \in \mathbb{R}$ is the reliance value at time step $n$, $s \in [0,1]$ is the reliance inertia, and $\epsilon$ is a sequence of i.i.d zero-mean Gaussian random variables as noise. The \emph{belief of autonomy capability}, $B_C(n) \in \mathbb{R}$ is derived as:
\begin{equation} \label{eq:baseBelief}
\begin{aligned}
    &B_C(n) = \\
        &\begin{cases}
            B_C(n-1) + {b_1}\left(C(n-1) - B_C(n-1)\right),&\text{if } I_{C} = 1\\
            B_C(n-1) + {b_0}(B_{C_{\text{ini}}} - B_C(n-1)),&\text{if } I_{C} = 0\\
        \end{cases}
\end{aligned}
\end{equation}
where $I_C \in {0, 1}$ defines different autonomy types by describing when capability information is available to the operator, $b_1 \in [0,1]$ describes the system interface transparency, and $b_0 \in [0,1]$ describes to what degree the operator's initial belief matters when system capability information is not available (i.e. when $I_C = 0$). Lastly, $C(n) \in \mathbb{R}^{\geq 0}$, and $C_\text{ini} \in \mathbb{R}^{\geq 0}$, are the true capability of the autonomy given the task, and its initial value, respectively. The value of $C(n)$ is defined in a later section. In this work, we will be focusing on situations with $I_C = 1$, meaning that the operator has access to the aid's capability information at all time steps (e.g., through some form of interface).

\begin{remark} \label{rem:realistic_capability}
One may incorrectly assume that to increase the number of reliance decisions made by a human operator, we need only to artificially increase the capability value provided to the reliance model. However, doing so would certainly violate the assumptions and reasoning behind the chosen reliance model. As we design for \emph{appropriate reliance}, modifying the true capability information of the autonomy would undermine that goal. 
\end{remark}

We now begin to extend existing work by modifying the reliance model given in \eqref{eq:baseBelief} to include an \emph{agreement} bias in the autonomy capability belief portion of the model. While this particular modification is certainly novel, the idea that agreement with an autonomous aid can sway human opinions is known \cite{losey2019robots,beck2014effect}. This modification models situations wherein the human operator's opinion of the aid is swayed by the ADA confirming the operator's initial selection. The agreement-adjusted belief is as follows:
\begin{equation} \label{eq:agreementBelief}
    \begin{aligned}
    B_C(n) &= B_C(n-1) + {b_1}\left(C(n-1) - B_C(n-1)\right)\\ 
    &\hspace{5mm}+ {A_C}{b_2}\left( 1 - B_C(n-1) \right)
    \end{aligned}
\end{equation}
where $b_2 \in [0,1]$ describes the degree to which aid agreement affects the operator's belief in autonomy capability, and $A_C \in \{-1, 1\}$ indicates whether the operator and ADA are in agreement ($A_C = -1$ if in disagreement). The agreement adjusted model in \eqref{eq:agreementBelief} remains based on DFT and incorporates the agreement bias into the operator's belief of capability, in keeping with the rest of the reliance model.

Following \cite{dubois2020adaptive}, preference is converted to a reliance state $d \in \left\{0,1\right\}$, where $d = 1$ indicates the operator is relying upon the ADA to select a task and $d = 0$ indicates the operator is selecting a task manually, by comparing the preference value with a threshold value $\theta \in \mathbb{R}^+$. Specifically, if $P(n) < \theta$ at time step $n$ then $d$ is set to $0$ and if $P(n) \geq \theta$ then $d$ is set to $1$. In this way we convert an internal continuous preference value to a concrete reliance decision by the operator, and $d$ can then be used later to inform the indicator model.

Note that the reliance model described above may not apply to all operators. As shown by \cite{akash2018classification,gao2006extending}, there is significant variance in how an individual interacts with a decision aid that depends on many personal characteristics. As such the parameters (see Table \ref{tab:parameters}) are assumed to be randomly selected from a corresponding distribution. Further, not all possible realizations from the parameter set would result in a reasonable reliance model, however given the relatively tight ranges mentioned in \cite{gao2006extending} we can assume that the chosen parameter distributions generate a set of reliance models that is a super set of realistic human reliance models. For completeness, all model parameters are described and defined in Table~\ref{tab:parameters}.
\begin{table}[t]
    \small
    \centering
    \caption{Variable Descriptions}
    \begin{tabular}{|c|l|}
        \hline
        $b_1 \in [0,1]$ & System interface transparency weighting \\
        \hline
        $b_2 \in [0,1]$ & ADA agreement weighting  \\
        \hline
        $s \in [0,1]$ & Controls reliance inertia \\
        \hline
        $C(n) \in \mathbb{R}^{\geq 0}$ & True ADA capability \\
        \hline
        $I_C \in \left\{ 0, 1 \right\}$ & Capability information indicator \\
        \hline
        $A_C \in \left\{ -1, 1 \right\}$ & Operator agreement indicator\\
        \hline
        $B_C(n) \in \mathbb{R}$ & Operator belief over capability\\
        \hline
        $P(n) \in \mathbb{R}$ & Operator preference value \\
        \hline
        $\theta \in \mathbb{R}^{\geq 0}$ & Operator reliance threshold \\
        \hline
        $d \in \left\{ 0, 1 \right\}$ & Operator reliance state\\
        \hline
        $S_i$ & Task set available at time step $i$ \\
        \hline
        $h_i \in S_i$ & Initial task selection\\ 
        \hline
        $a_i \in S_i$ & Decision aid task suggestion/advice\\
        \hline
        $a_{\text{opt}} \in S_i$ & Optimal task in expectation\\
        \hline
        $h^*_i \in S_i$ & Final task selection\\
        \hline
        $K \in \mathbb{R}^{\geq 0}$ & Total number of time steps \\
        \hline
        $V_i \in \mathbb{R}^L$ & Environmental context vector at time step $i$ \\
        \hline
        $R_E(\cdot) \in \mathbb{R}$ & Reward function, context dependent \\
        \hline
        $P_{\text{ind}}(n) \in \mathbb{R}$ & Indicator model preference value \\
        \hline
        $d_{\text{ind}} \in \left\{ 0, 1 \right\}$ & Indicator model reliance state\\
        \hline
    \end{tabular}
    \label{tab:parameters}
\end{table}

\subsection{Problem Statement} \label{sub:problem_statement}
\begin{problem} \label{prb:main_problem}

Given a human operator, applying a reliance model $P(n)$ with unknown stochastic parameters, interacting with an autonomous decision aid which results in a selected task $h_i^* \in S_i$, an environmental context vector $V_i$, multi-agent task set $S_i$, and a reward function $R_E(\cdot)$, approximately optimize the total reward received over $K$ iterations:
\begin{equation} \label{eq:main_problem}
    S^* = \underset{a_i \in S_i}{\text{argmax}} \sum_{i = 0}^K R_E \left( h_i^*, V_i \right)
\end{equation}
where $h_i^*$ is selected subject to the operator reliance model $P(n)$, and $R_E(h_i^*, V_i)$ is a reward function based on the selected task and current environment state as described by the context vector. The result of solving \eqref{eq:main_problem} is a set of suggestions, $S^*$, for each time step to approximately optimize total reward received. Note that the ADA does \emph{not} have direct control over the final task selected, $h_i^*$, thus selecting what task to suggest is the only method of control.
\end{problem}

\section{Decision Aid Structure} \label{sec:decision_aid_structure}
Here we detail the structure of the ADA and all of its component parts, which references the system flow diagram in Figure~\ref{fig:system_diagram} as well as the game structure discussed in Section~\ref{sub:game_structure}. The ADA can begin once the operator has made an initial selection, $h_i$, which in our case is derived from the $2018$ choice prediction competition (CPC18) data set \cite{bourgin2019cognitive} (detailed in the next subsection). The next concern is whether the operator will be relying upon the ADA or not, which is the purpose of the indicator model. Given $h_i$, the indicator model determines which suggestion the ADA makes, if reliance is indicated then the numerically optimal task is suggested, if reliance is \emph{not} indicated then the task predictor is used to make the suggestion. Using the task predictor at the specified times allows the ADA to take advantage of the agreement bias inherent in \eqref{eq:agreementBelief} to improve the overall reward received over $K$ iterations. In our considered problem, improvement is certainly possible as the human operators often do not select the optimal task due a variety of factors, such as the reward received in the previous trial biasing their reasoning \cite{bourgin2019cognitive}.

\begin{remark} \label{rem:reliance_controller}
Depending upon the application, attempting to match the operator's selection at every iteration may not always be the best path forward for an ADA. Indeed if the operator were using the ADA in an unsafe fashion, it may be advantageous to attempt to \emph{decrease} reliance where appropriate. While our formulation certainly allows this functionality, due to limitations of the selected data set and human testing requirements, we leave it to future work.
\end{remark}

\subsection{Human Decision-Making Data Set} \label{sub:dataset}

To both validate and train our predictor model we require a source of human decision data under some form of risk. To this end, consider the $2018$ choice prediction competition (CPC18) data set \cite{bourgin2019cognitive}, which tabulates data from a set of participants, sourced from Amazon Mechanical Turk, making sequential selection decisions with stochastic rewards based on the selection. This data set is uniquely suited to our problem due to its sequential nature, a rare property in human decision-making data sets, which allows us to implement the predictor portion of the ADA. Participants in the CPC18 data set played a set of $30$ \emph{games}, each game consisting of $25$ sequential decisions called trials. Each game consists of two gambles, each with their own reward distribution, where the participant was asked to make a selection between the two. After each trial the participants were presented with the reward gained/lost as well as the foregone reward. Each participant played $30$ different games drawn from a population of $270$ different paired distributions. A total of $926$ individuals participated, which generated $\approx 0.694\text{M}$ data points of human decision-making under reward-based risk. 

As an example of the contents of the CPC18 data set, a single game could be displayed as option "A" having a $25\%$ chance of generating $3$ reward and a $75\%$ chance of generating $0$ reward, and option "B" having a $20\%$ chance of generating $4$ reward and a $80\%$ chance of generating $0$. Typically, in the CPC18 data set option "B" is the more risky of the two. Readers are referred to \cite{plonsky2019predicting,bourgin2019cognitive} for more details on the data set and the construction of individual games.

To map this data set to our context of multi-agent task allocation, we first assume that each option represents a task or set of tasks in $S_i$. When the participant makes a selection we can interpret that the operator assigning a task to a single agent, and the reward generated is a function of the selected task. Going back to our running SAR example, each option in the data set might correspond to a method of search. One method covering a large area of land quickly, at a high risk of failing to locate a target, and the other method being a slower more methodical search, with a low risk of failing to locate a target.

\subsection{Task Predictor} \label{sub:predictor}
Given this data set, we can design and build a neural network to predict the next selection a human is likely to pick given the previous $k \in \mathbb{N}^+$ selections as well as a context vector. In our case, the context vector contains the game information from the data set, with rewards normalized to be within the interval $\left[ 0, 1 \right]$. We opted to use an LSTM-based network with $128$ recurrent neurons per layer, $4$ hidden layers, and a single dense output layer with a Softmax activation. LSTM layers were used, with an input size of $k$, as they are designed for use with sequence-to-sequence prediction problems \cite{hochreiter1997long}. The network was trained using randomized sequences drawn from CPC18, where the target was chosen to be the next selection by the participant.

\begin{figure}[t]
    \centering
    \includegraphics[width = 0.7\linewidth]{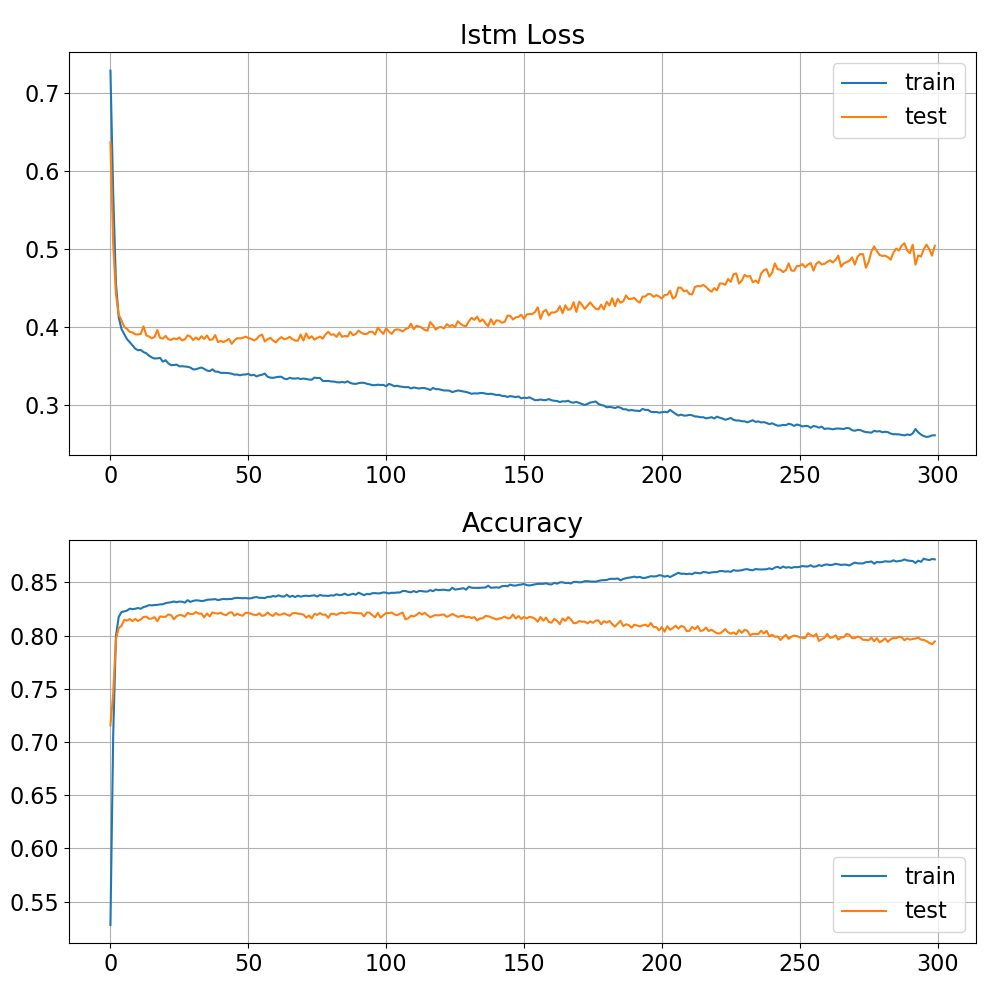}
    \caption{Showing the training and validation curves of predictor.}
    \label{fig:nn_training}
\end{figure}

The training was run for $75$ epochs, as this was seen to be the point at which performance stopped increasing and over-fitting began, see Figure~\ref{fig:nn_training} for illustration. After training the prediction accuracy in the validation set is $\approx 83\%$. Note that the prediction problem considered is quite difficult due to the natural variance between participants and their personal risk analysis, though well-suited to machine learning as creating a closed form model to predict human selections would be challenging. To the authors' knowledge this is the first time a sequential action predictor of this kind has been applied to the CPC18 data set and integrated into a human-autonomy teaming context.

\subsection{Autonomous Capability} \label{sub:automation_capability}
We can calculate the capability of the autonomy, given the context of the CPC18 data set, as the probability of selecting the gamble that generates more reward overall, see Section~\ref{sub:dataset} for details. Calculating capability is necessary to correctly implement the operator reliance model, this method also serves to determine the theoretical optimal task to select regardless of any interaction constraints. Given two binomial gamble reward random variables, $g_A$ and $g_B$, the capability of the autonomy, as we have posed it, is given by:
\begin{equation} \label{eq:automationCapability}
    \begin{aligned}
        &Pr(g_A < g_B) = \sum_{i = 0}^{n_c}\sum_{j = 0}^{i} P(g_B = i) P(g_A = j)\\
        &C(n) = Pr\left(g_A > \left\lceil \frac{n_c(L_B - L_A) + g_B(H_B - L_B}{H_A - L_A} \right\rceil \right)
    \end{aligned}
\end{equation}
where $g_A \sim \mathcal{B}(n_c, p_A)$ and $g_B \sim \mathcal{B}(n_c, p_B)$, $n_c$ is the number of selections left before the end of the game ($n_c = 25 - n$ in CPC18), $H_A,\,L_A,\,H_B$, and $L_B$ are the high and low, rewards from selecting option A and B respectively. We use $\mathcal{B}(n,p)$ to indicate a binomial distribution with $n$ samples and a probability $p$. The parameters controlling reward distributions form our environmental context vector $V_i$, here the size of the context vector would be $L = 6$. Here we are assuming, without loss of generality, that option A is the option more likely to generate maximum reward. If we evaluate $C(n)$ and find that it is $< 0.5$ we can simply take the opposite option and let $g_B$ be the maximum expected reward option. Let this optimal task be $a_{\text{opt}}$. Note that this formulation of capability is specific to the task options in the chosen data set, but could represent a variety of multi-agent task allocation scenarios.

\subsection{Indicator Model} \label{sub:indicator_model}
Certainly we do not have access to each operator's internal reliance model, thus we require a method of determining whether or not the operator is likely to take the ADA's suggestion in any given trial. With this capability, we can modulate suggestions in response to operator actions to improve overall team performance. Note that estimating whether the operator will rely upon the ADA is different from \emph{predicting} which task the operator would select.

We begin by assuming we have been given a \emph{disturbed} version of the parameters used in the reliance model from \eqref{eq:baseReliance}. This disturbed model can then be used as an indicator model for actual operator reliance decisions, we label it as $P_{\text{ind}}(n)$ along with its reliance state $d_{\text{ind}}$. To use the indicator model, we must provide the same inputs as for the real reliance model, such as capability and agreement, then we can use information from operator interactions to update the parameters. We use approximate Bayesian computation (ABC) \cite{lintusaari2017fundamentals} to update indicator model parameters as new interaction data becomes available. The mechanism of ABC is to compute a large set of realizations from a randomized model and compare it to observed data, through a set of summary statistics\footnote{We use the sample mean, variance, and skew as our summary statistics} and a Euclidean distance metric. Each realization that is within some threshold of the observed data is added to a set of accepted samples. The set of parameters used to generate each accepted sample become the priors for the parameters used to generate observed data.

The main advantage to using ABC, as opposed to for example a direct Bayesian or other common filtering paradigm, is that we retain freedom over all possible reliance models without the need to derive likelihoods based on the specific model in use. In addition, ABC provides convenient methods to update parameters iteratively as new data is observed but without recomputing large amounts of data. Readers are referred to \cite{lintusaari2017fundamentals,JMLR:v19:17-374} for more detailed information on ABC.

\subsection{Update Intervals} \label{sub:update_intervals}
Due to the fact that more interaction data becomes available as trials pass, we must periodically update the indicator model's parameters using ABC to increase accuracy. The interval to wait is certainly application specific, however in this work we use a period of $10$ complete game iterations, or $250$ individual trials, between updates. Selecting the update interval is a trade off between indicator model accuracy and computation time with diminishing returns for smaller values due to the lack of new observed data.

In addition, as previously mentioned, a large set of participants were used to obtain the data. Specifically in CPC18, every participant played $30$ game iterations. To reflect this fact in simulation, we reinitialize the reliance model every $30$ games with new realizations of stochastic parameters to simulate a new operator arriving. Of course refreshing the operator's parameters also means that we must reset the indicator model accordingly, otherwise the indicator model would be using inaccurate parameters to inform suggestions. The complete statement of the decision aid algorithm is given in Algorithm~\ref{alg:ADA_main}.

\begin{algorithm}[t]
  \caption{ADA Process}
  \label{alg:ADA_main}
  \begin{algorithmic}[1]
    \Procedure{Interact}{$D$}\Comment{Interact given data $D$}
        \State{$d_{\text{list}} = \{\}$}\Comment{Observed data for filter}
        \For{$(h_i,\, V_i) \text{ in } D$}
            \State{Predict selection given $V_i$}\Comment{Gives $\hat{h}_i$}
            \If{$d_{\text{ind}} = 1$}
                \State{$a_i\gets a_{\text{opt}}$}\Comment{Reliance is indicated}
            \Else
                \State{$a_i\gets \hat{h}_i$}\Comment{Reliance is \emph{not} indicated}
            \EndIf
            \State{$A_C \gets 1$ if $h_i = a_i$, $-1$ otherwise}
            \If{$d = 1$}
                \State{$h^*_i \gets a_i$}\Comment{Operator takes suggestion}
            \Else
                \State{$h^*_i \gets h_i$}\Comment{Operator ignores suggestion}
            \EndIf
            \State{$d_{\text{list}} = \{d_{\text{list}},\, (d,\, A_C)\}$}\Comment{Append new data}
            \State{Receive reward $R_E(h^*_i, V_i)$}
            \State{Step $P(n)$ and $P_{\text{ind}}(n)$}\Comment{$A_C$ used here}
            \If{Update required}
                \State{$P_{\text{ind}} \gets \text{ABC}(d_{\text{list}})$}\Comment{Periodic update}
            \EndIf
            \If{New operator}
                \State{Reinitialize both $P_{\text{ind}}(n)$ and $P(n)$}
            \EndIf
        \EndFor
    \EndProcedure
  \end{algorithmic}
\end{algorithm}

\section{Simulations} \label{sec:simulations}
\begin{table}[t]
    \small
    \centering
    \caption{Simulation Parameters}
    \begin{tabular}{|c|l|}
        \hline
        ABC samples & $10^4$ via rejection\\
        \hline
        ABC batch size & $10^5$\\
        \hline
        ABC threshold & $0.5$\\
        \hline
        Reset interval & $30$ iterations\\
        \hline
        $b_1$ distribution & $\mathcal{U}(0.01, 0.04)$\\
        \hline
        $b_2$ distribution & $\mathcal{U}(0.01, 0.04)$\\
        \hline
        $s$ distribution & $\mathcal{U}(0.10, 0.80)$\\
        \hline
        $\theta$ distribution & $\mathcal{U}(0.50, 0.20)$\\
        \hline
    \end{tabular}
    \label{tab:sim_parameters}
\end{table}
\begin{figure*}[t]
\centering
\subfloat[Myopic Case]{\includegraphics[width=0.475\linewidth]{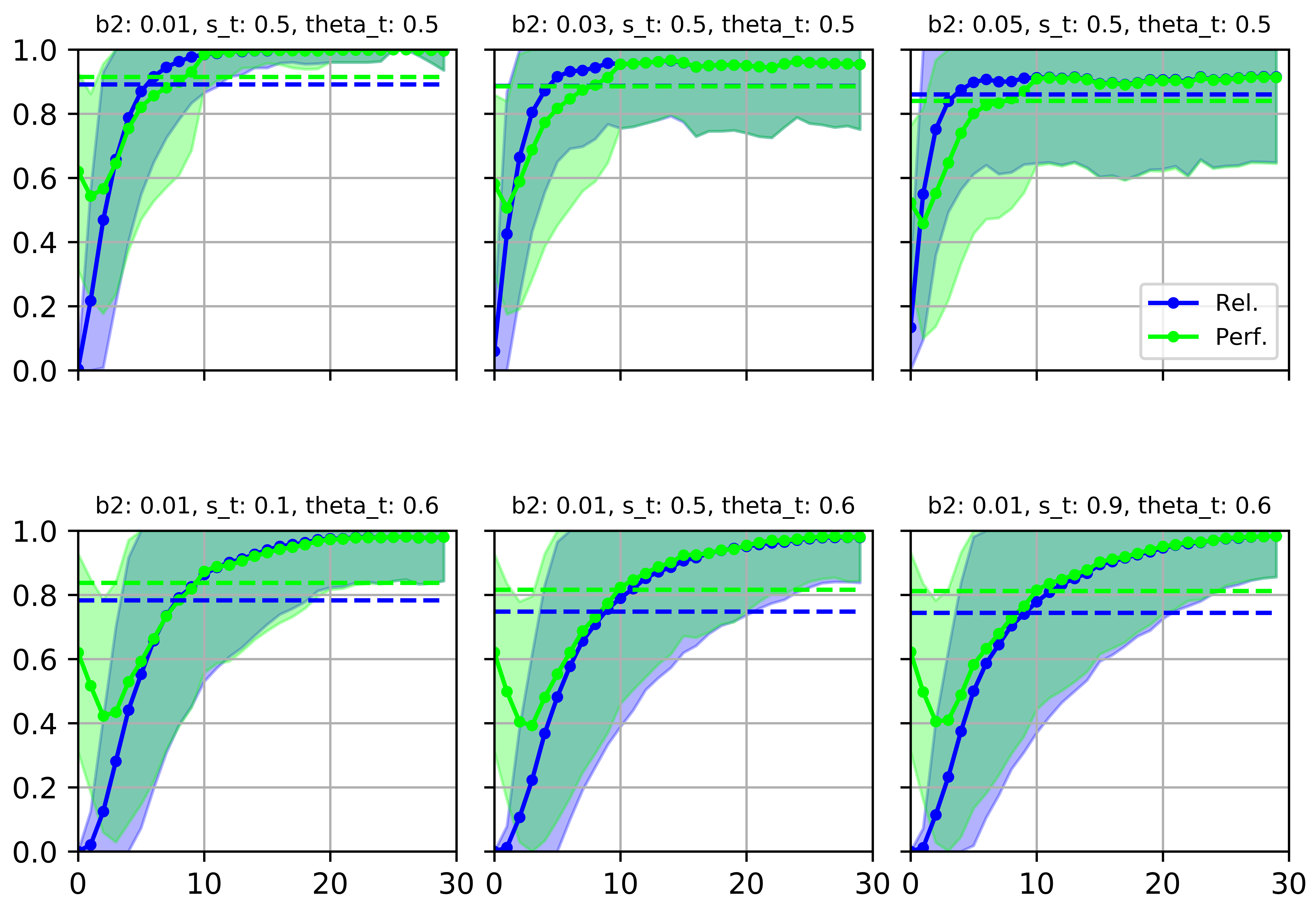}
\label{fig:main_sim_naive}}
\centering
\hfil
\subfloat[Predictive Case]{\includegraphics[width=0.475\linewidth]{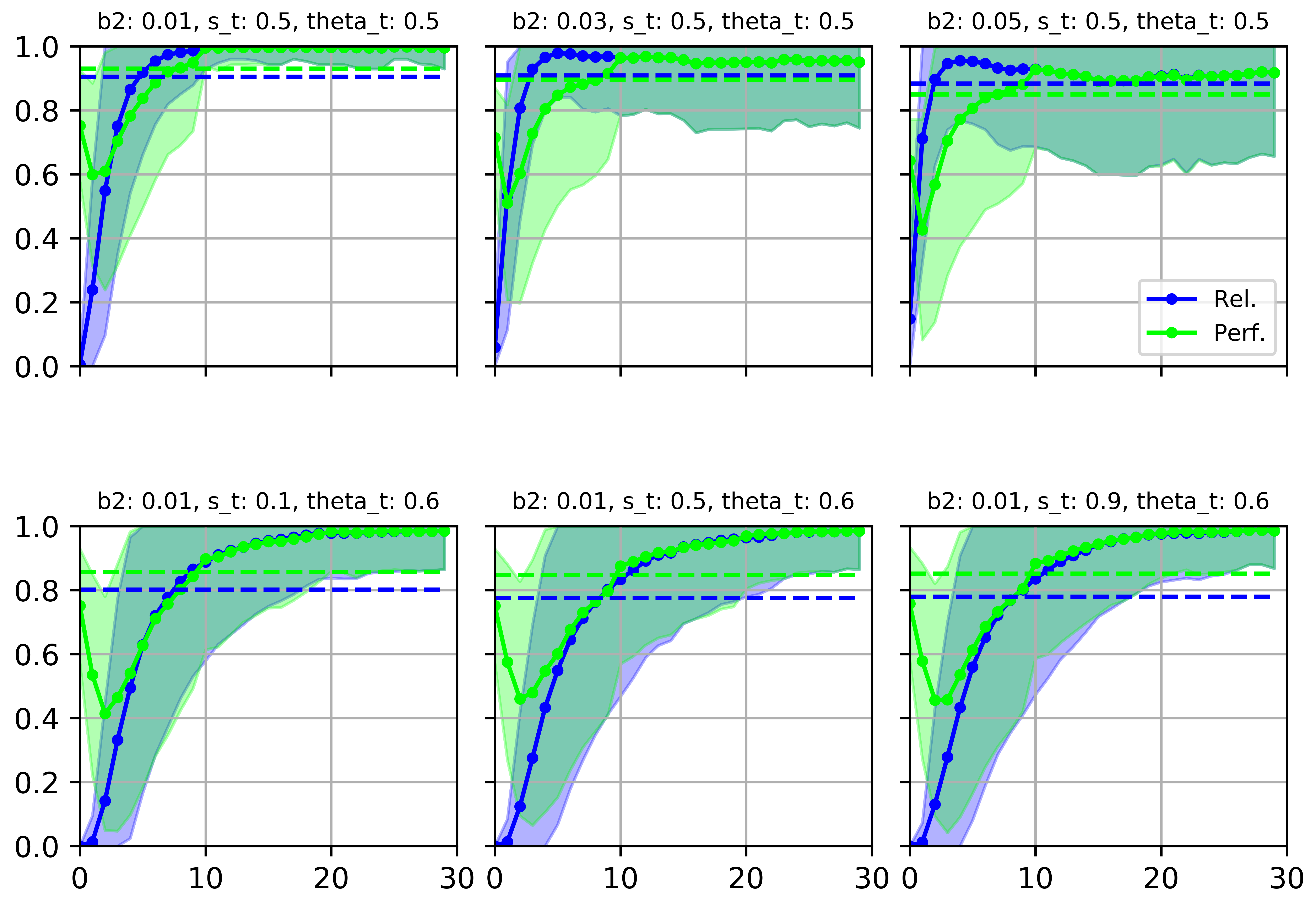}
\label{fig:main_sim_full}}
\caption{Showing a set of Monte Carlo simulations where each point corresponds to the mean reliance/performance observed during a complete game where each new operator is randomly initialized according to the parameters at the top of each plot. The blue and green lines represent reliance ($d$) and performance ($\rho$) respectively, and the shaded regions indicate $\pm1$ standard deviation. The dashed lines represent the mean of their corresponding value. The x-axis is $[0,30]$ since this is the number of games each participant played in the CPC18 data set.}
\label{fig:main_sims}
\end{figure*}

\subsection{Implementation} \label{sub:implementation}
The simulation pipeline was implemented in Python 3.7.7 and makes use of several common open source packages such as NumPy and PyTorch, the complete codebase used to generate results can be found \href{https://git.caslab.ece.vt.edu/hlarkin3/reliancecontroller}{here}\footnote{Link text: https://git.caslab.ece.vt.edu/hlarkin3/reliancecontroller}. Prior to simulation, the task predictor was trained on a portion of the CPC18 data set which resulted in a prediction accuracy of $\approx 80\%$ in the validation set. The training set is kept entirely separate from the data used in the simulation pipeline to prevent bias in the predictor's accuracy. The training process must be done beforehand due to the time required, as is common in machine learning applications.

In addition to the predictor, the indicator model parameter filter was implemented using an engine for likelihood-free inference (ELFI) \cite{JMLR:v19:17-374} packaged for use with Python. There are many different variations on the standard idea of ABC, however in this work we select rejection sampling, discussed in Section~\ref{sub:indicator_model}, as it is well suited to our case of periodically updating the indicator model. Other sampling methods are certainly applicable, sequential Monte Carlo sampling \cite{lintusaari2017fundamentals} was also considered and found to be unnecessary given the relative simplicity of the reliance model considered and frequency of updates.

In the interest of examining behaviors from a large set of reliance models simultaneously, each operator's reliance model parameters are randomized according to the distributions in Table~\ref{tab:sim_parameters}. Note the distributions used are uniform distributions of the form $\mathcal{U}(a,b)$ where $b$ is the distribution width. The type and width of each distribution was inspired by values used in \cite{gao2006extending} as well as by experimentation. Note that a reliance model based on \eqref{eq:agreementBelief} will have significantly different behaviors with small changes in its parameters. For example, decreasing the preference inertia value, $s$, will result in an operator that tends to switch reliance states often. Similarly, increasing the reliance threshold value, $\theta$, results in less overall reliance due to the high value of preference, from \eqref{eq:baseReliance}, required. We claim that the parameters and ranges selected are expressive enough to represent a wide range of human operators. 

Later in this section, we will be comparing the proposed method, including the task predictor and indicator model, to the case of myopic suggestions. Myopic here meaning that the decision aid \emph{only} suggests the numerically optimal task $a_{\text{opt}}$ at every iteration regardless of the indicator model. Comparing to the myopic case allows us to better quantify the potential benefits of the task predictor and indicator model system, as opposed to the widely used advice-only decision aid \cite{okamura2020empirical,akash2018classification,beck2004effects}. 

In addition to reliance as an indicator of success, we will also be using indicator model and prediction \emph{performance} as a metric, which we label $\rho(n)$. Specifically, we use the following metric for performance:
\begin{equation}\label{eq:ind_performance}
\rho(n) = 
\begin{cases}
1 &\text{ if } d_{\text{ind}} = d = 1\\
1 &\text{ if } d_{\text{ind}} = d = 0 \text{ and } a_i = h_i \\
0 &\text{ otherwise}\\
\end{cases}
\end{equation}
where $\rho(n)$ is evaluated at each time step $n$ and quantifies the joint indicator model and task selection prediction accuracy. Essentially, if the indicator model is correct \emph{and}, in the case of non-reliance, the ADA makes a correct prediction, then $\rho(n) = 1$. We do not allow $\rho(n) = 1$ in the case of $d_{\text{ind}} \neq d$, nor $d_{\text{ind}} = d = 0$ with $a_i \neq h_i$, since while the indicator model may be correct the suggested task is not. Taking the mean across all iterations in the data set allows us to examine the mean accuracy of the ADA.

\subsection{Results} \label{sub:sim_results}
\begin{table*}
\caption{Method Comparison}\label{tab:comparison_table}
\centering
\begin{tabular}{c c c}
$\theta = 0.5$ & $\theta = 0.6$ & $\theta = 0.7$  \\ 
{\begin{tabular}{|c|c|c|c|}
\hline
\backslashbox{$s$}{$b_2$} & 0.01 & 0.03 & 0.05 \\ \hline
0.1 & 4.14 & 3.19 & 4.52 \\ \hline
0.5 & 7.56 & 5.83 & 5.57 \\ \hline
0.9 & 7.21 & 7.84 & 6.98 \\ \hline
\end{tabular}}
&
{\begin{tabular}{|c|c|c|c|}
\hline
\backslashbox{$s$}{$b_2$} & 0.01 & 0.03 & 0.05 \\ \hline
0.1 & 6.96 & 11.34 & 8.86 \\ \hline
0.5 & 8.96 & 7.27 & 8.02 \\ \hline
0.9 & 11.35 & 11.62 & 10.61 \\ \hline
\end{tabular}}
&
{\begin{tabular}{|c|c|c|c|}
\hline
\backslashbox{$s$}{$b_2$} & 0.01 & 0.03 & 0.05 \\ \hline
0.1 & 25.01 & 20.68 & 12.50 \\ \hline
0.5 & 19.77 & 27.75 & 28.28 \\ \hline
0.9 & 7.98 & 15.21 & 10.50 \\ \hline
\end{tabular}}
\end{tabular}
\end{table*}
\subsubsection{Qualitative Results} \label{ssub:qualitative_results}
Here we present results of simulating the proposed method using the CPC18 data set. Shown in Figure~\ref{fig:main_sims} is the mean reliance and performance values across the data set for the two cases. At the top of each plot is the parameter distribution information used to randomize each new operator reliance model, where the value shown is the center of a uniform distribution of width $0.005$. We limit the stochasticity of parameters in this way to better examine the expected effect of the ADA in different circumstances.

The performance, from \eqref{eq:ind_performance}, is plotted in Figures \ref{fig:main_sim_full} and \ref{fig:main_sim_naive} which shows the average indicator model and predictor accuracy as human-autonomy interaction proceeds. There is a clear convergence towards the reliance graph as the indicator model increases in accuracy, the predictor accuracy does not improve over time after the first $k = 5$ interactions as this is the input length of the predictor. In certain plots a clear jump in performance after $10$ games is seen due to the ABC parameter update, where we are using prior interaction data to improve the indicator model. A smaller jump can also be seen after $20$ games as the second ABC update occurs, but the refinement is minimal by this point.

Along the top row of Figure~\ref{fig:main_sim_naive}, the $b_2$ parameter distribution center increases from left to right demonstrating the effect of different ranges of the agreement bias parameter. Here demonstrating once again that even slight parameter changes can generate significantly different operator profiles. Note that with higher values of $b_2$ the reliance tends to reach a lower value overall, this is likely due to the predictor's inaccuracies being weighted higher than other factors such as capability. Along the bottom row of Figure~\ref{fig:main_sim_naive} we instead vary the $s$ parameter, as expected we observe a relatively slight delay in convergence with higher values of $s$ but without much change to the shape. Further, the bottom row uses $\theta = 0.6$ and the result is a much decreased rate of reliance increase.

Moving on to Figure~\ref{fig:main_sim_full}, where in this case we are using the method described in Algorithm~\ref{alg:ADA_main} to generate suggestions. The same parameter ranges as in Figure~\ref{fig:main_sim_naive} are used to permit direct comparisons. Most notably, the speed of reliance increase is higher in the plotted cases, indicating the value of providing carefully selected suggestions. As a result the mean reliance achieved in the predictive case is higher in most cases shown. In addition, with higher ranges of $b_2$ there is some amount of reliance overshoot, caused by inertia present in \eqref{eq:baseReliance} and a high weight placed upon predictions compared to autonomy capability.

\subsubsection{Quantitative Results} \label{ssub:quantitative_results}

Shown in Table~\ref{tab:comparison_table} are results of comparing the previously mentioned myopic method to the predictive case. Specifically, in each cell we take the percentage difference between the overall mean reliance values of the two methods, where the reliance models are being driven by the parameter distribution centers shown on the left and top of the tables. Here we are examining the reliance value achieved over $30$ games and taking the mean across the participants, which corresponds to the dashed blue lines in Figure~\ref{fig:main_sims}. With Table~\ref{tab:comparison_table}, we are able to test a wide set of parameter ranges and see that the proposed method improves significantly upon the myopic method in nearly all cases tested, with some exceptions where the percentage improvement is $\leq 5\%$ and considered negligible.  

Interestingly in Table~\ref{tab:comparison_table}, as $\theta$ increases the margin of improvement seems to increase as well. In general terms, a higher preference threshold value results in an operator much less likely to rely on autonomy for task selection, thus it is encouraging to see the effectiveness of the proposed method in such cases. There are a few cases which dispute that trend however, such as $\theta = 0.7$ with a high $s$ value, which may be due to the limited number of trials used. That is, with higher $\theta$ reliance typically takes longer to build up, an effect worsened by a high inertia parameter as well.

\section{Conclusions} \label{sec:conclusions}
In this work we developed a game-theoretic modeling of the interaction between a human operator and an ADA in a multi-agent task allocation setting. In this setting, the ADA as designed to correctly calibrate the operator's reliance upon the aid and improve long-term team performance. We approached this via a combination of human action prediction and parameter fitting, resulting in a decision aid that adapts to human reliance dynamics. The proposed method was tested against a large set of simulated operators, and shown to substantially improve human-autonomy interactions compared to a myopic decision aid. 

\bibliographystyle{ieeetran.bst}
\bibliography{arxiv.bib}

\begin{thebibliography}{10}
\providecommand{\url}[1]{#1}
\csname url@samestyle\endcsname
\providecommand{\newblock}{\relax}
\providecommand{\bibinfo}[2]{#2}
\providecommand{\BIBentrySTDinterwordspacing}{\spaceskip=0pt\relax}
\providecommand{\BIBentryALTinterwordstretchfactor}{4}
\providecommand{\BIBentryALTinterwordspacing}{\spaceskip=\fontdimen2\font plus
\BIBentryALTinterwordstretchfactor\fontdimen3\font minus
  \fontdimen4\font\relax}
\providecommand{\BIBforeignlanguage}[2]{{%
\expandafter\ifx\csname l@#1\endcsname\relax
\typeout{** WARNING: IEEEtran.bst: No hyphenation pattern has been}%
\typeout{** loaded for the language `#1'. Using the pattern for}%
\typeout{** the default language instead.}%
\else
\language=\csname l@#1\endcsname
\fi
#2}}
\providecommand{\BIBdecl}{\relax}
\BIBdecl

\bibitem{music2017control}
S.~Musi{\'c} and S.~Hirche, ``Control sharing in human-robot team
  interaction,'' \emph{Annual Reviews in Control}, vol.~44, pp. 342--354, 2017.

\bibitem{okamura2020empirical}
K.~Okamura and S.~Yamada, ``Empirical evaluations of framework for adaptive
  trust calibration in human-ai cooperation,'' \emph{IEEE Access}, vol.~8, pp.
  220\,335--220\,351, 2020.

\bibitem{lee2004trust}
J.~D. Lee and K.~A. See, ``Trust in automation: Designing for appropriate
  reliance,'' \emph{Human factors}, vol.~46, no.~1, pp. 50--80, 2004.

\bibitem{gao2006extending}
J.~Gao and J.~D. Lee, ``Extending the decision field theory to model operators'
  reliance on automation in supervisory control situations,'' \emph{IEEE
  Transactions on Systems, Man, and Cybernetics-Part A: Systems and Humans},
  vol.~36, no.~5, pp. 943--959, 2006.

\bibitem{chen2020trust}
M.~Chen, S.~Nikolaidis, H.~Soh, D.~Hsu, and S.~Srinivasa, ``Trust-aware
  decision making for human-robot collaboration: Model learning and planning,''
  \emph{ACM Transactions on Human-Robot Interaction (THRI)}, vol.~9, no.~2, pp.
  1--23, 2020.

\bibitem{losey2019robots}
D.~P. Losey and D.~Sadigh, ``Robots that take advantage of human trust,''
  \emph{arXiv preprint arXiv:1909.05777}, 2019.

\bibitem{burton2020systematic}
J.~W. Burton, M.-K. Stein, and T.~B. Jensen, ``A systematic review of algorithm
  aversion in augmented decision making,'' \emph{Journal of Behavioral Decision
  Making}, vol.~33, no.~2, pp. 220--239, 2020.

\bibitem{Williams2020-sv}
R.~K. Williams, N.~Abaid, J.~McClure, N.~Lau, L.~Heintzman, A.~Hashimoto,
  T.~Wang, C.~Patnayak, and A.~Kumar, ``Collaborative multi-robot multi-human
  teams in search and rescue,'' \emph{Proceedings of the International ISCRAM
  Conference}, vol.~17, Apr. 2020.

\bibitem{Heintzman2021-mw}
L.~Heintzman, A.~Hashimoto, N.~Abaid, and R.~K. Williams, ``Anticipatory
  planning and dynamic lost person models for {Human-Robot} search and
  rescue,'' in \emph{2021 {IEEE} International Conference on Robotics and
  Automation ({ICRA})}.\hskip 1em plus 0.5em minus 0.4em\relax
  ieeexplore.ieee.org, May 2021, pp. 8252--8258.

\bibitem{hashimoto2019agent}
A.~Hashimoto and N.~Abaid, ``An agent-based model of lost person dynamics for
  enabling wilderness search and rescue,'' in \emph{Dynamic Systems and Control
  Conference}, vol. 59155.\hskip 1em plus 0.5em minus 0.4em\relax American
  Society of Mechanical Engineers, 2019, p. V002T13A005.

\bibitem{xu2008usariem}
X.~Xu, M.~Amin, and W.~Santee, ``Usariem technical report t08-05: Probability
  of survival decision aid (psda),'' 2008.

\bibitem{Heintzman2020-cz}
L.~Heintzman and R.~K. Williams, ``Nonlinear observability of unicycle
  multi-robot teams subject to nonuniform environmental disturbances,''
  \emph{Auton. Robots}, vol.~44, no.~7, pp. 1149--1166, Sep. 2020.

\bibitem{Heintzman2021-qr}
------, ``Multi-agent intermittent interaction planning via sequential greedy
  selections over position samples,'' \emph{IEEE Robot. Autom. Lett.}, vol.~6,
  no.~2, pp. 534--541, Apr. 2021.

\bibitem{koester2008lost}
R.~J. Koester, \emph{Lost Person Behavior: A search and rescue guide on where
  to look - for land, air and water}.\hskip 1em plus 0.5em minus 0.4em\relax
  dbs Productions LLC, 2008.

\bibitem{nam2019models}
C.~Nam, P.~Walker, H.~Li, M.~Lewis, and K.~Sycara, ``Models of trust in human
  control of swarms with varied levels of autonomy,'' \emph{IEEE Transactions
  on Human-Machine Systems}, vol.~50, no.~3, pp. 194--204, 2019.

\bibitem{wang2014human}
Y.~Wang, Z.~Shi, C.~Wang, and F.~Zhang, ``Human-robot mutual trust in (semi)
  autonomous underwater robots,'' in \emph{Cooperative Robots and Sensor
  Networks 2014}.\hskip 1em plus 0.5em minus 0.4em\relax Springer, 2014, pp.
  115--137.

\bibitem{Williams2017-ca}
R.~K. Williams, A.~Gasparri, and G.~Ulivi, ``Decentralized matroid optimization
  for topology constraints in multi-robot allocation problems,'' \emph{2017
  IEEE International Conference on Robotics and Automation}, 2017.

\bibitem{liu2019submodular}
J.~Liu and R.~K. Williams, ``Submodular optimization for coupled task
  allocation and intermittent deployment problems,'' \emph{IEEE Robotics and
  Automation Letters}, vol.~4, no.~4, pp. 3169--3176, 2019.

\bibitem{Sung2020-tl}
Y.~Sung, A.~K. Budhiraja, R.~K. Williams, and P.~Tokekar, ``Distributed
  assignment with limited communication for multi-robot multi-target
  tracking,'' \emph{Auton. Robots}, 2020.

\bibitem{Liu2021-gw}
J.~Liu, L.~Zhou, P.~Tokekar, and R.~K. Williams, ``Distributed resilient
  submodular action selection in adversarial environments,'' \emph{IEEE
  Robotics and Automation Letters}, vol.~6, no.~3, pp. 5832--5839, Jul. 2021.

\bibitem{Sung2018-mm}
Y.~Sung, A.~K. Budhiraja, R.~K. Williams, and P.~Tokekar, ``Distributed
  simultaneous action and target assignment for {Multi-Robot} {Multi-Target}
  tracking,'' in \emph{2018 {IEEE} International Conference on Robotics and
  Automation ({ICRA})}, May 2018, pp. 1--9.

\bibitem{liu2020data}
J.~Liu and R.~K. Williams, ``Data-driven models with expert influence: A hybrid
  approach to spatiotemporal process estimation,'' in \emph{Proceedings of the
  {IEEE/RSJ} International Conference on Intelligent Robots and Systems}, 2020,
  pp. 2467--2473.

\bibitem{liu2020coupled}
------, ``Coupled temporal and spatial environment monitoring for multi-agent
  teams in precision farming,'' in \emph{Proceedings of the IEEE Conference on
  Control Technology and Applications}, 2020, pp. 273--278.

\bibitem{liu2018optimal}
------, ``Optimal intermittent deployment and sensor selection for
  environmental sensing with multi-robot teams,'' in \emph{Proceedings of the
  IEEE International Conference on Robotics and Automation}, 2018, pp.
  1078--1083.

\bibitem{liu2020monitoring}
------, ``Monitoring over the long term: Intermittent deployment and sensing
  strategies for multi-robot teams,'' in \emph{Proceedings of the IEEE
  International Conference on Robotics and Automation}, 2020, pp. 7733--7739.

\bibitem{akash2018classification}
K.~Akash, W.-L. Hu, N.~Jain, and T.~Reid, ``A classification model for sensing
  human trust in machines using eeg and gsr,'' \emph{ACM Transactions on
  Interactive Intelligent Systems (TiiS)}, vol.~8, no.~4, pp. 1--20, 2018.

\bibitem{dubois2020adaptive}
C.~Dubois and J.~Le~Ny, ``Adaptive task allocation in human-machine teams with
  trust and workload cognitive models,'' in \emph{2020 IEEE International
  Conference on Systems, Man, and Cybernetics (SMC)}.\hskip 1em plus 0.5em
  minus 0.4em\relax IEEE, 2020, pp. 3241--3246.

\bibitem{beck2014effect}
G.~M. Beck, R.~Limor, V.~Arunachalam, and P.~R. Wheeler, ``The effect of
  changes in decision aid bias on learning: Evidence of functional fixation,''
  \emph{Journal of Information Systems}, vol.~28, no.~1, pp. 19--42, 2014.

\bibitem{bourgin2019cognitive}
D.~D. Bourgin, J.~C. Peterson, D.~Reichman, S.~J. Russell, and T.~L. Griffiths,
  ``Cognitive model priors for predicting human decisions,'' in
  \emph{International conference on machine learning}.\hskip 1em plus 0.5em
  minus 0.4em\relax PMLR, 2019, pp. 5133--5141.

\bibitem{plonsky2019predicting}
O.~Plonsky, R.~Apel, E.~Ert, M.~Tennenholtz, D.~Bourgin, J.~C. Peterson,
  D.~Reichman, T.~L. Griffiths, S.~J. Russell, E.~C. Carter \emph{et~al.},
  ``Predicting human decisions with behavioral theories and machine learning,''
  \emph{arXiv preprint arXiv:1904.06866}, 2019.

\bibitem{hochreiter1997long}
S.~Hochreiter and J.~Schmidhuber, ``Long short-term memory,'' \emph{Neural
  computation}, vol.~9, no.~8, pp. 1735--1780, 1997.

\bibitem{lintusaari2017fundamentals}
J.~Lintusaari, M.~U. Gutmann, R.~Dutta, S.~Kaski, and J.~Corander,
  ``Fundamentals and recent developments in approximate bayesian computation,''
  \emph{Systematic biology}, vol.~66, no.~1, pp. e66--e82, 2017.

\bibitem{JMLR:v19:17-374}
\BIBentryALTinterwordspacing
J.~Lintusaari, H.~Vuollekoski, A.~Kangasr{\"a}{\"a}si{\"o}, K.~Skyt{\'e}n,
  M.~J{\"a}rvenp{\"a}{\"a}, P.~Marttinen, M.~U. Gutmann, A.~Vehtari,
  J.~Corander, and S.~Kaski, ``Elfi: Engine for likelihood-free inference,''
  \emph{Journal of Machine Learning Research}, vol.~19, no.~16, pp. 1--7, 2018.
  [Online]. Available: \url{http://jmlr.org/papers/v19/17-374.html}
\BIBentrySTDinterwordspacing

\bibitem{beck2004effects}
G.~M. Beck, \emph{The effects of decision aid bias and fixation on user
  performance}.\hskip 1em plus 0.5em minus 0.4em\relax University of
  Missouri-Columbia, 2004.

\end{thebibliography}

\end{document}